\documentclass[a4paper,12pt]{article}
\usepackage[pctex32]{graphics}

\begin{document}
\newcommand{\cl}{\centerline}
\renewcommand{\theequation}{\arabic{equation}}
\newcommand{\beq}{\begin{equation}}
\newcommand{\eeq}{\end{equation}}
\newcommand{\bea}{\begin{eqnarray}}
\newcommand{\eea}{\end{eqnarray}}
\newcommand{\nn}{\nonumber}
\newcommand\pa{\partial}
\newcommand\un{\underline}
\newcommand\ti{\tilde}
\newcommand\pr{\prime}
\begin{titlepage}
\setlength{\textwidth}{5.0in} \setlength{\textheight}{7.5in}
\setlength{\parskip}{0.0in} \setlength{\baselineskip}{18.2pt}

\begin{flushright}
{\tt gr-qc/0605084} \\
{\tt SOGANG-HEP 317/06}
\end{flushright}

\vspace*{0.05cm}

\begin{center}
{\large\bf Entropy of the Randall-Sundrum brane world with the
generalized uncertainty principle}
\end{center}

\begin{center}
{Wontae Kim$^{1,a}$, Yong-Wan Kim$^{2,b}$, and Young-Jai
Park$^{1,c}$} \par
\end{center}

\begin{center}
{$^{1}$Department of Physics and Center for Quantum Spacetime,}\par
{Sogang University, Seoul 121-742, Korea}\par \vskip 0.2cm
{$^{2}$National Creative Research Initiative Center for Controlling
Optical Chaos, Pai-Chai University, Daejeon 302-735, Korea}\par
\end{center}
\vskip 0.05cm
\begin{center}
(\today)
\end{center}
\vskip 0.05cm
\begin{center}
{\bf ABSTRACT}
\end{center}
\begin{quotation}

By introducing the generalized uncertainty principle, we calculate
the entropy of the bulk scalar field on the Randall-Sundrum brane
background without any cutoff. We obtain the entropy of the massive
scalar field proportional to the horizon area. Here, we observe that
the mass contribution to the entropy exists in contrast to all
previous results of the usual black hole cases with the generalized
uncertainty principle.

\vskip 0.1cm



\vskip 0.05cm

\noindent
$^a$wtkim@sogang.ac.kr \\
\noindent $^b$ywkim@pcu.ac.kr \\
\noindent $^c$yjpark@sogang.ac.kr
\noindent

\end{quotation}
\end{titlepage}

\newpage

\section{Introduction}

Three decades ago, Bekenstein had suggested that the entropy of a
black hole is proportional to the area of the horizon through the
thermodynamic analogy \cite{bek}. Subsequently, Hawking showed
that the entropy of the Schwarzschild black hole satisfies exactly
the area law by means of Hawking radiation based on the quantum
field theory \cite{haw}. After their works, 't Hooft investigated
the statistical properties of a scalar field outside the horizon
of a Schwarzschild black hole by using the brick wall method with
the Heisenberg uncertainty principle \cite{tho}. However, although
he obtained the entropy proportional to the horizon area, an
unnatural brick wall cutoff was introduced to remove the
ultraviolet divergence near horizon \cite{gm,kkps,acb,kop,med,lz}.
A similar idea was also considered by the entanglement entropy
interpretation with a momentum cutoff, which is almostly
equivalent to the brick wall model \cite{bkl}. Recently, many
efforts \cite{gup} have been devoted to the generalized
uncertainty relations, and its consequences, especially the effect
on the density of states. Very recently, in Refs.
\cite{li,liu,kkp}, the authors calculated the entropy of black
holes by using the novel equation of state of density motivated by
the generalized uncertainty principle \cite{gup}, which
drastically solves the ultrviolet divergences of the just vicinity
near the horizon without any cuttoff.

On the other hand, much interests have been paid to the Randall and
Sundrum model to resolve the gauge hierarchy problem \cite{rs1,rs2},
which is based upon the fact that our universe may be embedded in
higher-dimensional spacetimes \cite{add}. Furthermore, various
aspects \cite{vars} of this model have been studied including the
cosmological applications \cite{cos}. To study quantum mechanical
aspect of this black brane world, we may first consider its entropy,
which is expected to satisfy the area law \cite{haw}. However, up to
now, the statistical entropy of the black brane world was only
studied by using the brick wall method \cite{acb,kop,med,lz} with
the Heisenberg uncertainty principle, which still has several
difficulties including artificial ultraviolet and infrared cutoffs.

In this paper, we would like to study the entropy of a bulk scalar
field on the black brane background avoiding the difficulty in
solving the 5-dimensional Klein-Gordon wave equation by using the
quantum statistical method. By using the novel equation of state
of density motivated by the generalized uncertainty principle in
the quantum gravity \cite{li,liu,kkp}, we derive the free energy
of a massive scalar field through the complete decomposition of
the extra mode and the radial mode, which has been impossible in
the brick wall method. We then calculate the quantum entropy of
the black hole via thermodynamic relation between the free energy
and the entropy. As a result, we obtain the desired entropy
proportional to the horizon area without any artificial cutoff and
little mass approximation. Here, we newly observe that in contrast
to all previous results \cite{li,liu,kkp} of the usual black hole
cases with the generalized uncertainty principle, which is
independent of the ordinary scalar field mass, the contribution of
the bulk scalar mass to the entropy for the Randall and Sundrum
brane case exists.

\section{Scalar field on Randall and Sundrum brane Background}

Let us start with the following action \cite{rs1,rs2} of the Randall
and Sundrum model in $(4+1)$ dimensions,
\begin{eqnarray}
 \label{2baction}
 S_{(5)} &=& \frac{1}{16 \pi G^{(5)}_N} \int d^4 x \int dy \sqrt{-g_{(5)}}
            \left[ R^{(5)} + 12k^2 \right] \nn \\
         &{}&- \int d^4 x \left[\sqrt{-g_{(+)}} \lambda_{(+)} +
\sqrt{-g_{(-)}} \lambda_{(-)} \right],
\end{eqnarray}
where $\lambda_{(+)}$ and $\lambda_{(-)}$ are tensions of the
branes at $y=0$ and $y=\pi r_c$, respectively, while $12k^2$ is a
cosmological constant. We assume that orbifold $S^1/Z_{2}$
possesses a periodicity in the extra coordinate $y$, and identify
$-y$ with $y$. Two singular points on the orbifold are located at
$y=0$ and $y=\pi r_c$. Two 3-branes are placed at these points.
Note that the metric on each brane is defined as $ g_{\mu
\nu}^{(+)} \equiv g_{\mu\nu}^{(5)}(x^{\mu},y=0)$ and $g_{\mu
  \nu}^{(-)} \equiv g_{\mu\nu}^{(5)}(x^{\mu}, y=\pi r_c)$.

Since we are interested in black holes, let us assume the bulk
metric as
\begin{equation}
   \label{metrican}
  ds^2_{(5)}=e^{-2ky\Theta(x)} g_{\mu\nu}(x) dx^{\mu}dx^{\nu} + T^2(x)dy^2,
\end{equation}
where $T(x)$ is the moduli field. Note that we use $\mu, \nu = 0, 1,
\cdots, 3$ for brane indices. Among possible solutions satisfying
the equations of motion \cite{kop}, let us consider the
4-dimensional Schwarzschild black hole solution as a slice of AdS
spacetime as a brane solution,
\begin{equation}
  ds^2 = e^{-2ky}\left[ - \left(1-\frac{2M}{r}\right) dt^2
        + \left(1-\frac{2M}{r}\right)^{-1} dr^2
        + r^2 d\Omega^2_{(2)} \right] + dy^2,
 \label{d4ss}
\end{equation}
where $d\Omega^2_{(2)}$ is a metric of unit 2-sphere and we set
$G_{(4)} = 1$ for convenience. In fact, it is a black string
solution intersecting the brane world, which describes a black hole
placed on the hypersurface at the fixed extra coordinate.

In this brane background, let us first consider a bulk scalar field
with mass $m$, which satisfies the Klein-Gordon equation,
\begin{equation}
  \label{wveqn}
  (\nabla^2_{(5)} - m^2)\Phi = 0,
\end{equation}
which is explicitly given as
\begin{eqnarray}
  \label{wveqn2}
  & & e^{2ky}\left[ -\frac{1}{f} \partial_{t}^2 \Phi + \frac{1}{r^2}
  \partial_{r}\left(r^2 f \partial_{r} \Phi\right) + \frac{1}{r^2 {\rm
  sin}\theta } \partial_{\theta} ({\rm sin} \theta \partial_{\theta}
  \Phi) + \frac{1}{r^2 {\rm sin}^2 \theta} \partial_{\phi}^2 \Phi\right]
  \nonumber \\
  & & + e^{4ky} \partial_{y}(e^{-4ky} \partial_{y} \Phi) - m^2 \Phi = 0,
\end{eqnarray}
where $f = 1- \frac{2M}{r}$. If we set
\begin{equation}
\label{5d}
 e^{4ky}  \partial_{y}(e^{-4ky} \partial_{y} \chi) -
m^2 \chi  + \mu^2 e^{2ky}\chi = 0
\end{equation}
with $ \Phi(t,r,\theta, \phi, y) \equiv \Psi(t,r,\theta,
\phi)\chi(y)$, then the separation of variables is easily done, and
the reduced form of the effective field equation becomes
\begin{equation}
  \label{eqn2}
   -\frac{1}{f} \partial_{t}^2 \Psi + \frac{1}{r^2} \partial_{r}
   \left(r^2 f \partial_{r} \Psi \right) + \frac{1}{r^2 {\rm sin} \theta}
   \partial_{\theta} ({\rm sin}\theta \partial_{\theta} \Psi)
   + \frac{1}{r^2 {\rm sin}^2 \theta} \partial_{\phi}^2 \Psi - \mu^2 \Psi = 0.
\end{equation}
Note that the above eigenvalue $\mu^2$ plays a role of the effective
mass on the brane. Substituting the 4-dimensional wave function
$\Psi(t,r,\theta, \phi) = e^{-i\omega t}\psi(r, {\theta}, \phi)$, we
find that the 4-dimensional Klein-Gordon equation becomes
\begin{equation}
\label{rtheta0} \partial_{r}^2 \psi + \left( \frac{1}{f}
\partial_{r} f + \frac{2}{r}\right)
\partial_{r} \psi + \frac{1}{f}
\left({\frac{1}{r^2}}\left[\partial^2_\theta + {\rm cot}\theta
\partial_\theta + {\frac{1}{{\rm sin}^{2}\theta}}\partial^2_\phi \right] +
\frac{\omega^{2}}{f} - \mu^{2} \right)\psi = 0.
\end{equation}
By using the Wenzel-Kramers-Brillouin (WKB) approximation \cite{tho}
with $\psi \sim exp[iS(r,\theta,\phi)]$, we have
\begin{equation}
\label{wkb} {p_{r}}^{2} = \frac{1}{f}\left(\frac{\omega^{2}}{f} -
\mu^{2} - \frac{p^2_\theta}{r^2} - \frac{p_{\phi}^2}{r^2 {\rm sin}^2
\theta} \right),
\end{equation}
where
\begin{equation}
\label{mom} p_{r} = \frac{\partial S}{\partial r},~
 p_{\theta} = \frac{\partial S}{\partial \theta},
 ~p_{\phi} = \frac{\partial S}{\partial \phi}.
\end{equation}

Furthermore, we also obtain the square module momentum as follows
\begin{equation}
\label{smom} p^{2} = p_{i}p^{i} = g^{rr}{p_{r}}^{2} + g^{\theta
\theta}{p_{\theta}}^{2}+ g^{\phi\phi}{p_{\phi}}^{2} =
\frac{\omega^{2}}{f} - \mu^{2}.
\end{equation}
Then, the volume in the momentum phase space is given by
\begin{eqnarray}
V_{p}(r,\theta)&=& \int dp_{r}dp_{\theta}dp_{\phi} \nonumber \\
&=& \frac{4\pi}{3} \sqrt{ \frac{1}{f}(\frac{\omega^2}{f}-\mu^2)}
\cdot \sqrt{r^2(\frac{\omega^2}{f}-\mu^2)} \cdot \sqrt{{r^2 {\rm
sin}^2 \theta}(\frac{\omega^{2}}{f} - \mu^{2})} \nonumber
\\&=&\frac{4\pi}{3} \frac{r^2 {\rm sin}\theta}{\sqrt{f}}
\left(\frac{\omega^2}{f}-\mu^2 \right)^{\frac{3}{2}}
\end{eqnarray}
with the condition $\omega\geq\mu\sqrt{f}$.

\section{Mode Spectrum}

Recently, many efforts have been devoted to the generalized
uncertainty relation \cite{gup} given by
\begin{equation} {\Delta x} {\Delta p} \ge \frac{\hbar}{2}\left(1 +
{\lambda}(\frac{{\Delta p}}{\hbar})^{2}\right).
\end{equation}
From now on, we take the units $\hbar=k_{B}=c \equiv 1$. Then, since
one can easily get ${\Delta x} \geq \sqrt{\lambda}$, which gives the
lowest bound, it can be defined to be the the minimal length near
horizon, which effectively plays a role of the brick wall cutoff.
Furthermore, based on the generalized uncertainty relation, the
3-dimensional volume of a phase cell is changed from $(2{\pi})^{3}$
into
\begin{equation}
\label{gup} (2{\pi})^{3}(1 + {\lambda}{p^{2}})^{3},
\end{equation}
where $p^2 = p^{i}p_{i}~(i = r, \theta, \phi).$

From the Eqs. (\ref{smom}) and (\ref{gup}), the number of  quantum
states related to the radial mode with energy less than $\omega$
is given by
\begin{eqnarray}
\label{Tnqs} n_{r}(\omega) &=& \frac{1}{(2\pi)^3} \int dr d\theta
d\phi dp_{r} dp_{\theta}dp_{\phi} \frac{1}{\left(1+
{\lambda}(\frac{{\omega}^2}{f}- \mu^{2})\right)^3} \nonumber   \\
&=&\frac{1}{(2\pi)^3} \int dr d\theta d\phi \frac{1}{\left(1+
{\lambda}\left(\frac{{\omega}^2}{f}- \mu^{2}\right) \right)^3} V_{p}(r,\theta) \nonumber   \\
&=& \frac{2}{3\pi}\int_{r_{H}} dr \frac{r^2}{\sqrt{f}}
\frac{\left(\frac{{\omega}^2}{f}-
\mu^{2}\right)^{\frac{3}{2}}}{\left(1+
{\lambda}(\frac{{\omega}^2}{f}- \mu^{2})\right)^3}.
\end{eqnarray}
It is interesting to note that $n_{r}(\omega)$ is convergent at the
horizon without any artificial cutoff due to the existence of the
suppressing $\lambda$-term in the denominator induced from the
generalized uncertainty principle.

On the other hand, the exact quantization of Eq.(\ref{5d}) seems to
be cumbersome. However, since in the WKB approximation \cite{tho},
the wave number $k_{\chi}$ of the wave function $\chi(y)$ is easily
written as
\begin{equation}
  \label{waveno1}
  k_{\chi}^2(y, \mu) = \mu^2 e^{2ky} - m^2,
\end{equation}
the number of extra mode for a given value $\mu$ is given by
\begin{equation}
  \label{modesE1}
 \pi n_{\chi}(\mu) = \int_{0}^{\pi r_{c}} dy~
 \sqrt{\mu^2 e^{2ky} - m^2}.
\end{equation}
We then obtain the total number of extra mode with energy less
than $\omega$ as follows
\begin{eqnarray}
  \label{modes11}
 \pi n_{\chi} &=& \int_{m}^{\frac{\omega}{\sqrt{f}}} d\mu~
 \frac{dn_{\chi}(\mu)}{d\mu}  \nonumber \\
 &=& \frac{1}{k}\int_{m}^{\frac{\omega}{\sqrt{f}}} d\mu
 \frac{1}{\mu} \left(\sqrt{\mu^2 e^{2k\pi r_{c}}-m^2}
 - \sqrt{\mu^2-m^2}\right).
\end{eqnarray}
As a result, we could formally write down the proper total number of
quantum states with energy less than $\omega$ as follows
\begin{equation}
\label{Ndef} N_{T}(\omega) = \int dN_{T}(\omega) = \int
~dn_{r}~dn_{\chi}.
\end{equation}

\section{Free Energy and Entropy}

For the bosonic case, the free energy at inverse temperature $\beta$
is given by
\begin{equation}
\label{def} e^{-\beta F} = \prod_K
                \left[ 1 - e^{-\beta \omega_K} \right]^{-1}~,
\end{equation}
where $K$ represents the set of quantum numbers. Then, by using
Eq. (\ref{Tnqs}), we are able to obtain the equation of free
energy as
\begin{eqnarray}
\label{TfreeE}
 F_{T}&=& \frac{1}{\beta}\sum_K \ln \left[ 1 - e^{-\beta \omega_K} \right]
   ~\approx ~\frac{1}{\beta} \int dN_{T}(\omega) ~\ln
            \left[ 1 - e^{-\beta \omega} \right]  \nonumber   \\
 &=& - \int^{\infty}_{\mu\sqrt{f}} d\omega
 \frac{N_T(\omega)}{e^{\beta\omega} -1} \nonumber  \\
   &=& - \frac{2}{3\pi} \int_{r_{H}} dr \frac{r^2}{\sqrt{f}}
   \int^{\infty}_{\mu\sqrt{f}} d\omega~\int_{m}^{\frac{\omega}{\sqrt{f}}} d\mu~
    \frac{\left(\frac{{\omega}^2}{f}
   - \mu^{2}\right)^{\frac{3}{2}}}{(e^{\beta \omega} -1)\left(1+ \lambda
   (\frac{{\omega}^2}{f}-\mu^2)\right)^3}
     \left(\frac{dn_{\chi}}{d\mu}\right) \nonumber  \\
   &=& - \frac{2}{3\pi} \int_{r_{H}} dr \frac{r^2}{\sqrt{f}}~ \Lambda_{T}
   \int_{m}^{\frac{\omega}{\sqrt{f}}} d\mu ~ \left(\frac{dn_{\chi}}{d\mu}\right).
\end{eqnarray}
Here, we have taken the continuum limit in the first line and
integrated by parts in the second line in Eq. (\ref{TfreeE}). Note
that our free energy is exactly the same as Eq. (27) in Ref.
{\cite{med}} with the exception of the new suppressing
$\lambda$-term that is introduced by the generalized uncertainty
principle in the denominator, which drastically solves the
ultraviolet divergence near the event horizon. In the last line,
since $f \rightarrow 0$ near the event horizon, {\it i.e.}, in the
range of $(r_H, r_H +\epsilon)$, $\frac{{\omega}^2}{f}
   - \mu^{2}$ becomes $\frac{{\omega}^2}{f}$.
Although we do not require the little mass approximation, the
integral equation of $\omega$ is naturally reduced to
\begin{equation}
\label{Tentropy1} \Lambda_{T} = \int^{\infty}_{0} d\omega
\frac{{f}^{-\frac{3}{2}}{\omega}^3}{(e^{\beta \omega} -1)\left(1+
\lambda
   \frac{{\omega}^2}{f}\right)^3}. \nonumber \\
\end{equation}
Therefore, the free energy can be rewritten as
\begin{equation}
\label{TfreeEf}
 F_{T}= - \frac{2}{3\pi} \int_{r_{H}} dr \frac{r^2}{f^2} \int^{\infty}_{0} d\omega
\frac{{\omega}^3}{(e^{\beta \omega} -1)\left(1+ \lambda
\frac{{\omega}^2}{f}\right)^3}\int_{m}^{\frac{\omega}{\sqrt{f}}}
d\mu \left(\frac{dn_{\chi}}{d\mu}\right).
\end{equation}
Now, since at this stage the $n_{\chi}$ mode part is completely
decoupled with the $n_{r}$ mode part for the $\mu$ coupling on the
contrary to the previous results \cite{kop,med}, we are able to
separately carry out the integral equation of $\mu$ in Eq.
(\ref{modes11}). As a result, we obtain
\begin{eqnarray}
\label{Extra} k \pi n_\chi &=& \left( \sqrt{\frac{\omega^2 e^{2k\pi
r_c}}{f}-m^2} -\sqrt{\frac{\omega^2}{f}-m^2} \right)- m \gamma \nonumber\\
&-& m \left(\tan^{-1}\sqrt{\frac{\omega^2e^{2k\pi r_c}}{m^2f}-1} -
\tan^{-1}\sqrt{\frac{\omega^2}{m^2f}-1}\right)
\end{eqnarray}
with $\alpha_a = e^{ak\pi r_c}-1 ~(a=1,2)$ and $\gamma =
\sqrt{\alpha_2} - \tan^{-1}\sqrt{\alpha_2} \geq 0$. Furthermore,
near the event horizon as $f\rightarrow 0$, we get the integral from
the Eq. (\ref{Extra}) without little mass approximation as
\begin{equation}
\label{modes12} \pi k n_\chi = \alpha_1 \frac{\omega}{\sqrt{f}} -
m\gamma.
\end{equation}
Hence, the value of free energy can be rewritten as
\begin{eqnarray}
\label{TfreeEf1}
 F_{T}= &-& \frac{2\alpha_1}{3\pi^{2}k} \int_{r_{H}} dr \frac{r^2}{f^2\sqrt{f}}
\int^{\infty}_{0} d\omega \frac{{\omega}^4}{(e^{\beta \omega}
-1)\left(1+ \lambda
\frac{{\omega}^2}{f}\right)^3} \nonumber \\
 &+& \frac{2m \gamma}{3\pi^{2}k} \int_{r_{H}} dr \frac{r^2}{f^2}  \int^{\infty}_{0}
d\omega \frac{{\omega}^{3}}{(e^{\beta \omega} -1)\left(1+ \lambda
\frac{{\omega}^2}{f}\right)^3}.
\end{eqnarray}
From this free energy and Eq. (\ref{modes12}), the entropy for the
scalar field is given by
\begin{eqnarray}
\label{Tentropy0} S_{T} &=& \beta^2 \frac{\partial F_T}{\partial
\beta} \nonumber \\
&=& \beta^2\frac{2\alpha_1}{3\pi^{2}k} \int_{r_{H}} dr
\frac{r^2}{f^2\sqrt{f}}  \int^{\infty}_{0} d\omega \frac{e^{\beta
\omega}{\omega^5}}{(e^{\beta\omega}-1)^2
(1+ \lambda \frac{\omega^2}{f})^3} \nonumber \\
&{}& - \beta^2 \frac{2m \gamma}{3\pi^{2}k} \int_{r_{H}} dr
\frac{r^2}{f^2} \int^{\infty}_{0} d\omega \frac{e^{\beta
\omega}{\omega}^{4}}{(e^{\beta \omega} -1)^2 \left(1+ \lambda
\frac{{\omega}^2}{f}\right)^3}
\nonumber \\
&\equiv& \frac{2 \alpha_1}{3\pi^{2}k} \int_{r_{H}} dr
\frac{r^2}{\sqrt{f}}~ \Lambda_{T}^1 - \frac{2m \gamma}{3\pi^{2}k}
\int_{r_{H}} dr \frac{r^2}{\sqrt{f}}~ \Lambda_{T}^2,
\end{eqnarray}
where
\begin{eqnarray}
\label{Gdef} \Lambda_{T}^1 &=& \int^\infty_0 dx \frac{f^{-2}
\beta^{-4}
x^{5}}{(1-e^{-x})(e^{x}-1)\left(1+\frac{\lambda}{\beta^{2}f}x^{2}\right)^3},\nonumber \\
\Lambda_{T}^2 &=& \int^\infty_0 dx \frac{f^{-\frac{3}{2}} \beta^{-3}
x^{4}}{(1-e^{-x})(e^{x}-1)\left(1+\frac{\lambda}{\beta^{2}f}x^{2}\right)^3}
\end{eqnarray}
with $x=\beta \omega$.

Now, let us rewrite Eq. (\ref{Tentropy0}) as
\begin{equation}
\label{Sdef} S_{T} = \frac{2\alpha_1}{3\pi^{2}k \lambda^2} ~
\int_{r_{H}} dr \frac{r^2}{\sqrt{f}}~ \Lambda_{T}^1 - \frac{2m
\gamma}{3\pi^{2}k \lambda^{\frac{3}{2}}} \int_{r_{H}} dr
\frac{r^2}{\sqrt{f}}~ \Lambda_{T}^2,
\end{equation}
where
\begin{eqnarray}
\label{Gdef1} \Lambda_{T}^1 &=& \int^\infty_0 dX
\frac{a^{2}X^{5}}{(e^{\frac{a}{2} X}-e^{-
\frac{a}{2}X})^{2}(1+X^2)^3}, \nonumber \\
 \Lambda_{T}^2 &=& \int^\infty_0 dX
\frac{a^{2}X^{4}}{(e^{\frac{a}{2} X}-e^{-
\frac{a}{2}X})^{2}(1+X^2)^3}
\end{eqnarray}
with $x=\beta\sqrt{\frac{f}{\lambda}}X \equiv aX$. Note that when
$r\rightarrow r_H$, $a$ goes to zero. Since we are interested in
the contributions from just the vicinity of the horizon, the
integrals in Eq. (\ref{Gdef}) are finally reduced as follows:
\begin{eqnarray}
\label{GI} \Lambda_{T}^1 = \int^\infty_0 dX \frac{X^3}{(1+X^2)^3}
=\frac{1}{4}, \nonumber \\
\Lambda_{T}^2 = \int^\infty_0 dX \frac{X^2}{(1+X^2)^3} =
\frac{\pi}{16}.
\end{eqnarray}

On the other hand, we are also interested in the contribution from
just the vicinity near the horizon in the range $(r_H, r_H +
\epsilon)$, where $\epsilon$ is related to a proper distance of
order of the minimal length, $\sqrt{\lambda}$ as follows
\begin{eqnarray}
\label{invariant} \sqrt{\lambda}=\int_{r_H}^{r_H +\epsilon}
                          \frac{dr}{\sqrt{f(r)}}
                \approx \int_{r_H}^{r_H +\epsilon}
                          \frac{dr}{\sqrt{2\kappa(r-r_{H})}}
                = \sqrt{\frac{2\epsilon}{\kappa}}.
\end{eqnarray}
Here $\kappa$ is the surface gravity at the horizon of the black
hole, and it is identified as $\kappa = 2\pi \beta$.

Therefore, when $r\rightarrow r_{H}$, we finally get the desired
entropy of the massive scalar field on the RS black brane
background as follows
\begin{eqnarray}
\label{finalS} S_{T} &\approx& \frac{2\alpha_1}{3\pi^{2}k \lambda^2}
\cdot r_H^2~ \sqrt{\lambda}\cdot \frac{1}{4} - \frac{2m
\gamma}{3\pi^{2}k
\lambda^{\frac{3}{2}}} \cdot r_H^2~ \sqrt{\lambda}\cdot \frac{\pi}{16} \nonumber \\
&=& \left(\frac{\alpha_1}{6\pi^{3}k}\frac{1}{(\sqrt{\lambda})^3} -
m \frac{ \gamma}{24 \pi^{2}k }\frac{1}{(\sqrt{\lambda})^2} \right)
~\frac{A}{4},
\end{eqnarray}
which is proportional to the area $A= 4 \pi r_{H}^2$. Note that
there is no divergence within the just vicinity near the horizon
due to the effect of the generalized uncertainty relation on the
quantum states.

It seems to be appropriate to comment on the entropy
(\ref{finalS}). First, by using the generalized uncertainty
principle, this entropy is obtained from the contribution of the
just vicinity near the horizon in the range $(r_H, r_H +
\epsilon)$, which is neglected by the brick wall method. Second,
since the entropy consists of the inverse power terms of the
minimal length, it is non-perturbative. Moreover, the positive
dominant leading term is proportional to $(\sqrt{\lambda})^{-3}$,
while the negative sub-leading term, which gives the massive
effect, is proportional to $(\sqrt{\lambda})^{-2}$.

In conclusion, we have investigated the massive bulk scalar field
within the just vicinity near the horizon of a static black hole in
the black brane world by using the generalized uncertainty
principle. We have derived the free energy of a massive scalar field
through the complete decomposition of the extra mode and the radial
mode, which has been impossible in the brick wall method with the
Heisenberg uncertainty principle. From this free energy, we have
obtained the desired entropy proportional to the two-dimensional
area of the black brane world without any artificial cutoff and
little mass approximation. As a result, we have newly observed that
the negative contribution of the bulk scalar mass to the entropy
exists for this brane case in contrast to all previous results,
which is independent of the mass of the ordinary scalar field, of
the usual black hole cases with the generalized uncertainty
principle.

\section*{Acknowledgments}

This work is supported by the Science Research Center Program of
the Korea Science and Engineering Foundation through the Center
for Quantum Spacetime of Sogang University with grant number
R11-2005-021. Y.-J. Park is also supported by the Korea Research
Foundation Grant funded by the Korean Government
(KRF-2005-015-C00105).

\end{document}